\documentclass[a4paper, article]{article} 

\usepackage{url}

\usepackage{graphics}
\usepackage{xcolor}
\usepackage[utf8]{inputenc}
\usepackage[T1]{fontenc}
\usepackage{amssymb} 
\usepackage{stmaryrd}

\usepackage{multicol} 
\usepackage[lofdepth,lotdepth]{subfig}
\usepackage{wrapfig}
\usepackage{tikz} 
\usetikzlibrary{arrows,positioning,automata,fit, decorations.pathreplacing, matrix}
\usetikzlibrary{shapes}
\usetikzlibrary{decorations.markings}
\usetikzlibrary{decorations.shapes}
\usetikzlibrary{arrows.meta}

\usepackage{caption}

\usepackage{amsmath} 
\usepackage{xspace} 

\usepackage{xparse} 
\usepackage{stmaryrd} 
\usepackage[babel=true]{csquotes} 

\usepackage[affil-it]{authblk} 

\usepackage{amsmath}

\usepackage{algorithmicx}
\usepackage[compatible]{algpseudocode}
\usepackage{algorithm}

\algtext*{ENDFOR}
\algtext*{ENDIF}
\algtext*{ENDMATCH}

\protected\def\NP{\ifmmode \mbox{\sc NP} \else {\sc NP}\xspace\fi}

\newcommand{\intervalproblem}{\textsf{PIC}\xspace} 
\newcommand{\setcover}{\textsf{SET COVER}\xspace}

\newcommand{\intervalpack}{P}
\newcommand{\intervalsolution}{\mathbb{S}}
\newcommand{\intervalproblemtarget}{N}
\newcommand{\intervalproblempacknumber}{M}

\newcommand{\problemdefinition}[2]{\begin{itemize}
\item Input: #1
\item Output: #2\end{itemize}}

\newcommand{\interval}[2]{\ensuremath{[#1, #2]}\xspace}

\newcommand{\genericinterval}{I}

\newcommand{\UNION}{\displaystyle\bigcup}				

\newcommand{\set}[1]{{\left\{#1\right\}}}

\newcommand{\explicationlitteral}[2]{\node at (-1, -#1) {\footnotesize #2};}
\newcommand{\explicationvariable}[2]{\node at (-1, -#1) {\footnotesize #2};}

\newcommand{\Set}[1]{\ensuremath{\{#1\}}\xspace}

\newcommand{\posinterval}[1]{\genericinterval^+_{#1}}
\newcommand{\neginterval}[1]{\genericinterval^-_{#1}}

\newcommand{\drawintervalle}[4]{
	\draw[#4] (#2, -#1) -- (#3,-#1);
	\foreach \x in {#2,..., #3} {
		\node[fill=#4,circle, inner sep=0.5mm] at (\x, -#1) {};
	}
}
\newcommand{\drawsingleton}[2]{
	\drawintervalle {#1} {#2} {#2} {red}
}	

\newcommand{\drawpack}[4]{\draw[rounded corners, draw=none, fill=gray!20!white] (#3-0.4, -#1) rectangle (#4+0.4, -#2);}

\newcommand{\drawpacksingleline}[3]{\drawpack{#1-0.3}{#1+0.3}{#2}{#3}}

\newcommand{\drawlineofnumbers}[2]{
	\foreach \x in {1, 2, 3, ..., #2} {
		\node (\x) at (\x, -#1) {\x};}
}

\newcommand{\drawgrid}[2]{
	\foreach \x in {1, 2, 3, ..., #2} {
		\draw[dotted, gray!50!white] (\x, 0) -- (\x, -#1);
	}
	
      }

\protected\def\SAT{\ifmmode \mbox{\sc {\textsf{SAT}}} \else {\sc {\textsf{SAT}}}\xspace\fi}
\newcommand{\clause}{\ensuremath{C}\xspace}
\newcommand{\literal}{\ensuremath{\ell}\xspace}

\newcommand{\threeSAT}{$3$-\SAT}
\newcommand{\threeBtwoSAT}{\textsf{($3$, B$2$)}-\SAT}
\newcommand{\threeBtwoSATclause}[3]{(#1 \lor #2 \lor #3)}
\newcommand{\threeBtwoSATprop}[1]{p_{#1}}
\newcommand{\threeBtwoSATliteral}[2]{\literal^{#1}_{#2}}
\newcommand{\threeBtwoSATvaluation}{\nu}
\newcommand{\genericsingleton}{S}
\newcommand{\genericsingletonbis}{S'}
\newcommand{\clausesingleton}[1]{{\mathit\genericsingleton}_{#1}}
\newcommand{\threeBtwoSATvariablepack}[1]{\mathit{VP}_{#1}}
\newcommand{\threeBtwoSATclausepack}{\mathit{CP}}
\newcommand{\threeBtwoSATpositiveclausepackone}[1]{\threeBtwoSATclausepack^{1+}_{#1}}
\newcommand{\threeBtwoSATpositiveclausepacktwo}[1]{\threeBtwoSATclausepack^{2+}_{#1}}
\newcommand{\threeBtwoSATnegativeclausepackone}[1]{\threeBtwoSATclausepack^{1-}_{#1}}
\newcommand{\threeBtwoSATnegativeclausepacktwo}[1]{\threeBtwoSATclausepack^{2-}_{#1}}

\newtheorem{theorem}{Theorem}

\makeatletter
\newcommand*{\algrule}[1][\algorithmicindent]{%
	\makebox[#1][l]{%
		\hspace*{.2em}
		\vrule height .75\baselineskip depth .25\baselineskip
	}
}

\newcount\ALG@printindent@tempcnta
\def\ALG@printindent{%
	\ifnum \theALG@nested>0
	\ifx\ALG@text\ALG@x@notext
	\else
	\unskip
	\ALG@printindent@tempcnta=1
	\loop
	\algrule[\csname ALG@ind@\the\ALG@printindent@tempcnta\endcsname]%
	\advance \ALG@printindent@tempcnta 1
	\ifnum \ALG@printindent@tempcnta<\numexpr\theALG@nested+1\relax
	\repeat
	\fi
	\fi
}
\patchcmd{\ALG@doentity}{\noindent\hskip\ALG@tlm}{\ALG@printindent}{}{\errmessage{failed to patch}}
\patchcmd{\ALG@doentity}{\item[]\nointerlineskip}{}{}{} 

\algrenewcommand\alglinenumber[1]{\tiny #1:}
\makeatother

\algrenewcommand\algorithmicindent{1mm}%

\makeatletter
\algrenewcommand\ALG@beginalgorithmic{\footnotesize}
\makeatother

\begin{document}

\title{The Packed Interval Covering Problem is \NP-complete}

\author{Abdallah Saffidine}
\affil{University of New South Wales, Sydney, Australia}

\author{Sébastien Lê Cong}
\affil{Univ Rennes, CNRS, IRISA France}

\author{Sophie Pinchinat}
  \affil{Univ Rennes, CNRS, IRISA France}  
  
\author{François Schwarzentruber}
\affil{Univ Rennes, CNRS, IRISA, France}


\maketitle
\pagebreak
We introduce a new decision problem, called Packed Interval Covering  (\intervalproblem),  defined as follows.


\problemdefinition
{an integer $\intervalproblemtarget > 0$ and a family of finite sets $\intervalpack_1$, \ldots, $\intervalpack_\intervalproblempacknumber$ (\emph{packs}) of subintervals of $\interval{1}{\intervalproblemtarget}$.}
{
  are there subintervals $\genericinterval_1 \in \intervalpack_1$, \ldots, $\genericinterval_\intervalproblempacknumber \in \intervalpack_\intervalproblempacknumber$ such that
  \[\interval{1}{\intervalproblemtarget} = \UNION_{k=1..\intervalproblempacknumber} \genericinterval_k?\]
}

\begin{figure}[h]
	\begin{center}
		\begin{tikzpicture}[scale=0.5]
		\explicationlitteral{1.8}{pack No 1}
		\explicationlitteral{3}{pack No 2}
		\explicationlitteral{4}{pack No 3}
		\drawlineofnumbers{-0.5}{9}
		
		\drawpack{2.2}{1.3}{1}{9}
		\drawpacksingleline{3}{1}{9}
		\drawpacksingleline{4}{1}{9}	
		\drawgrid{5}{9}
		\drawintervalle {0}{1}{9}{blue}
		\drawintervalle {1.5}{1}{6}{orange}
		\drawintervalle {1.9}{5}{9}{orange}
		\drawintervalle {3}{1}{3}{orange}
		\drawintervalle {3}{4}{6}{orange}
		\drawintervalle {3}{7}{7}{orange}
		
		\drawintervalle {4}{4}{4}{orange}
		\end{tikzpicture}
	\end{center}
	\caption{Example of an instance of the Packed Interval Covering Problem.\label{figure:picexample}}
\end{figure}
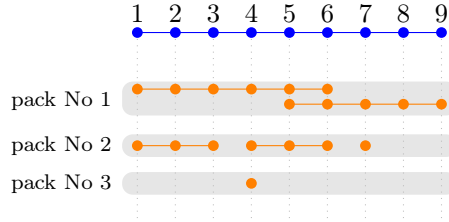

As a running example, consider 3 packs $\set{\interval{1}{6}, \interval{5}{9}}$, $\set{\interval{1}{3}, \interval{4}{6}, \interval{7}{7}}$, $\set{\interval{4}{4}}$. 
%
%
%
It is clear that we can cover $\interval{1}{9}$ by selecting $\interval{5}{9}$, $\interval{1}{3}$ and $\interval{4}{4}$ as shown in Figure~\ref{figure:picexample}.

Note that our \intervalproblem problem differs from the well-known \setcover \cite{DBLP:conf/coco/Karp72} on several aspects: first, in \intervalproblem, there are no quantities to minimize; second, \intervalproblem has a \emph{two-level} input (packs of subintervals) whereas \setcover has a one-level input (the collection of subsets is not organized in packs).

\begin{theorem}
	\label{thm:NP-hardnessPIC}
	 {\normalfont \intervalproblem} is \NP-complete, whether integers are encoded in unary or in binary.
\end{theorem}

The following non-deterministic algorithm establishes membership in NP: guess an interval in each pack, then check that the union of these intervals covers $\interval{1}{N}$.

For its \NP-hardness, we define a polynomial reduction from \threeBtwoSAT, which is the restriction of \threeSAT where each variable has exactly two positive occurrences and two negative occurrences.\footnote{As a side note, we could generalise the contruction below for \threeSAT or even \SAT, but resort to \threeBtwoSAT to avoid having to deal with some minor technicalities.} \threeBtwoSAT is \NP-complete~\cite{DBLP:journals/eccc/ECCC-TR03-049}.

%
	Let atoms $(\threeBtwoSATprop{i})_{1 \leq i \leq n}$ and clauses $(\clause_j)_{1 \leq j \leq m}$ be an instance of \threeBtwoSAT, with $\clause_j = \threeBtwoSATclause{\threeBtwoSATliteral{j}{i_1}}{\threeBtwoSATliteral{j}{i_2}}{\threeBtwoSATliteral{j}{i_3}}$ where $\threeBtwoSATliteral{j}{i} \in \Set{\threeBtwoSATprop{i}; \lnot \threeBtwoSATprop{i}}$. We build an instance of $\intervalproblem$ as follows, considering the target interval $\interval{1}{\intervalproblemtarget}$ where $\intervalproblemtarget = 4n+m$: the general idea lies in using each element from $\interval{4n+1}{4n+m}$ as a token that one specific clause is satisfied; a literal from the clause can switch the token on if and only if said literal evaluates to true. Meanwhile, $\interval{1}{4n}$ is used for the variables' truth values.

	To each variable $\threeBtwoSATprop{i}$, we associate a \emph{variable pack} $\threeBtwoSATvariablepack{i} = \Set{\posinterval{i}; \neginterval{i}}$ with $\posinterval{i} = \interval{4i-3}{4i-2}$ and $\neginterval{i} = \interval{4i-1}{4i}$. Notice that all such intervals are disjoint. Now, writing singleton $\clausesingleton{x} = \Set{4n+x}$ ($1 \leq x \leq m$), and $\clause_j$ and $\clause_k$ ($j < k$) the very two clauses where the positive literal $\threeBtwoSATprop{i}$ occurs, we define the two \emph{clause packs} $\threeBtwoSATpositiveclausepackone{i} = \Set{\Set{4i-3}; \clausesingleton{j}}$ and $\threeBtwoSATpositiveclausepacktwo{i} = \Set{\Set{4i-2}; \clausesingleton{k}}$. Likewise, clauses $\clause_l$ and $\clause_h$ ($l < h$) involving $\lnot \threeBtwoSATprop{i}$ yield the two clause packs $\threeBtwoSATnegativeclausepackone{i} = \Set{\Set{4i-1}; \clausesingleton{l}}$ and $\threeBtwoSATnegativeclausepacktwo{i} = \Set{\Set{4i}; \clausesingleton{h}}$.

\begin{figure}[h]
	\begin{center}
		\newcommand{\explicationsigne}[2]{\node at (#1, -1) {\footnotesize #2};}
		\newcommand{\explicationclause}[2]{\node at (#1, -1) {\footnotesize #2};}
		\newcommand{\nbvariablesthreeBtwoSATexample}{3}
		\newcommand{\nbclausesthreeBtwoSATexample}{4}
		\begin{tikzpicture}[xscale=0.5, yscale=0.4]
		\foreach \y in {2, ..., 16} {
				\drawpacksingleline{\y}{1}{16}	}
			
		\foreach \x in {1, 2, 3, ..., 16} {
			\node (\x) at (\x, 1) {\x};
			\draw[dotted, gray!50!white] (\x, -1) -- (\x, -16);
		}

		\foreach \y in {-2, ..., -16} {
			\draw[dotted, gray!20!white] (0, \y) -- (16, \y);
		}

		\drawintervalle 0 1 {16} {blue}
		
		\explicationvariable{2} {$\threeBtwoSATvariablepack{1}$}
		\drawintervalle {2} 1 {2} {orange}
		\drawintervalle {2} 3 {4} {orange}
		
		\explicationvariable{3} {$\threeBtwoSATvariablepack{2}$}
		\drawintervalle {3} 5 {6} {orange}
		\drawintervalle {3} 7 {8} {orange}
		
		\explicationvariable{4} {$\threeBtwoSATvariablepack{3}$}
		\drawintervalle {4} 9 {10} {orange}
		\drawintervalle {4} {11} {12} {orange}

		\foreach \i in {1, ..., \nbvariablesthreeBtwoSATexample} {
			\explicationsigne{4*\i - 2.5} {$\posinterval{\i}$}
			\explicationsigne{4*\i - 0.5} {$\neginterval{\i}$}
		}

		\foreach \i in {1, ..., \nbclausesthreeBtwoSATexample} {
			\explicationclause{4*\nbvariablesthreeBtwoSATexample + \i} {$\clausesingleton{\i}$}
		}
				
		\explicationlitteral{5} {$\threeBtwoSATpositiveclausepackone{1}$}
		\drawsingleton {5} {1}
		\drawsingleton {5} {13}
		\explicationlitteral{6} {$\threeBtwoSATpositiveclausepacktwo{1}$}
		\drawsingleton {6} {2}
		\drawsingleton {6} {14}
		\explicationlitteral{7} {$\threeBtwoSATnegativeclausepackone{1}$}
		\drawsingleton {7} {3}
		\drawsingleton {7} {15}
		\explicationlitteral{8} {$\threeBtwoSATnegativeclausepacktwo{1}$}
		\drawsingleton {8} {4}
		\drawsingleton {8} {16}
		\explicationlitteral{9} {$\threeBtwoSATpositiveclausepackone{2}$}
		\drawsingleton {9} {5}
		\drawsingleton {9} {13}
		\explicationlitteral{10} {$\threeBtwoSATpositiveclausepacktwo{2}$}
		\drawsingleton {10} {6}
		\drawsingleton {10} {16}
		\explicationlitteral{11} {$\threeBtwoSATnegativeclausepackone{2}$}
		\drawsingleton {11} {7}
		\drawsingleton {11} {14}
		\explicationlitteral{12} {$\threeBtwoSATnegativeclausepacktwo{2}$}
		\drawsingleton {12} {8}
		\drawsingleton {12} {15}
		\explicationlitteral{13} {$\threeBtwoSATpositiveclausepackone{3}$}
		\drawsingleton {13} {9}
		\drawsingleton {13} {13}
		\explicationlitteral{14} {$\threeBtwoSATpositiveclausepacktwo{3}$}
		\drawsingleton {14} {10}
		\drawsingleton {14} {16}
		\explicationlitteral{15} {$\threeBtwoSATnegativeclausepackone{3}$}
		\drawsingleton {15} {11}
		\drawsingleton {15} {14}
		\explicationlitteral{16} {$\threeBtwoSATnegativeclausepacktwo{3}$}
		\drawsingleton {16} {12}
		\drawsingleton {16} {15}

		\end{tikzpicture}
	\end{center}
	\caption{The instance of \intervalproblem associated to the example \threeBtwoSAT-instance ${\threeBtwoSATclause{\threeBtwoSATprop1}{\threeBtwoSATprop2}{\threeBtwoSATprop3} \land \threeBtwoSATclause{\threeBtwoSATprop1}{\lnot \threeBtwoSATprop2}{\lnot \threeBtwoSATprop3} \land 
			\threeBtwoSATclause{\lnot \threeBtwoSATprop1}{\lnot \threeBtwoSATprop2}{\lnot \threeBtwoSATprop3}
			\land 
			\threeBtwoSATclause{\lnot \threeBtwoSATprop1}{\threeBtwoSATprop2}{\threeBtwoSATprop3}}$.\label{figure:intervalselectionproblemnphardexample}
		}
              \end{figure}

	Let us suppose the \threeBtwoSAT instance positive and show that the \intervalproblem instance we built is positive too. From a valuation $\threeBtwoSATvaluation$ that satisfies the formula, we define an interval selection in the following way: whenever $\threeBtwoSATvaluation(\threeBtwoSATprop{i}) = 1$, we select $\posinterval{i}$ in $\threeBtwoSATvariablepack{i}$ and $\clausesingleton{j}$, $\clausesingleton{k}$, $\Set{4i-1}$ and $\Set{4i}$ in the four clause packs associated to $\threeBtwoSATprop{i}$; likewise, when $\threeBtwoSATvaluation(\threeBtwoSATprop{i}) = 0$, we select $\neginterval{i}$ in $\threeBtwoSATvariablepack{i}$ and $\clausesingleton{l}$, $\clausesingleton{h}$, $\Set{4i-3}$ and $\Set{4i-2}$ in the four clause packs associated to $\threeBtwoSATprop{i}$.

	Now, each element from $\interval{1}{4n}$ belongs to a selected interval: either that of a variable pack for some variable $\threeBtwoSATprop{i}$, or that of a clause pack for some clause where the literal for $\threeBtwoSATprop{i}$ evaluates to $0$. 
	Furthermore, each clause mapped to $1$---that is, all clauses---matches, by construction, a distinct selected singleton, and thus, $\interval{4n+1}{4n+m}$ is covered.

	\vspace{0.5em}

Conversely, we assume our PIC instance to be positive and show that the corresponding (3, B2)-SAT instance is positive as well.
        Firstly, we can turn any solution of \intervalproblem to one such that every integer in $\interval{1}{4n}$ is covered by exactly one interval; this is achievable by switching a clause-pack choice whenever the singleton is included in a selected interval from a variable pack.

	Using such a solution $\intervalsolution$, we can now derive a valuation $\threeBtwoSATvaluation$ as said above: namely, $\threeBtwoSATvaluation(\threeBtwoSATprop{i}) = 1$ iff $\posinterval{i}$ is selected in $\threeBtwoSATvariablepack{i}$. Let us consider a clause $\clause_x$; as $\intervalsolution$ is a solution, the element of $\clausesingleton{x}$ is covered. $\clausesingleton{x}$ is triggered by some literal for a variable $\threeBtwoSATprop{i}$ in $\clause_x$; thus, calling $\genericsingletonbis = \Set{y}$ the other singleton from $\clause_x$'s clause pack and observing that $y$ belongs to exactly one variable-pack interval and one clause-pack singleton, either $y \in \posinterval{i}$, which implies that $\posinterval{i}$ is selected and thus $\threeBtwoSATvaluation(\threeBtwoSATprop{i}) = 1$ while $\threeBtwoSATprop{i}$ appears as a positive literal in $\clause_x$ by construction, or $y \in \neginterval{i}$ and then $\threeBtwoSATvaluation(\threeBtwoSATprop{i}) = 0$ while $\threeBtwoSATprop{i}$ appears as a negative literal in $\clause_x$: in both cases, $\threeBtwoSATvaluation(\clause_x) = 1$. Consequently, $\threeBtwoSATvaluation$ satisfies all clauses.

Thus, the \threeBtwoSAT-instance is positive if and only if the corresponding \intervalproblem-instance is positive. This lets us conclude that \intervalproblem is NP-complete.


\paragraph{Historical note}

Sophie has addressed the problem \intervalproblem, Abdallah
has designed an elegant proof of its \NP-hardness, and Sébastien and François have worked on the writing of a pedagogical proof argument.

\paragraph{Acknowledgements}

We would like to thank Sébastien Chédor for offering his time in elaborating a first reduction for the hardness proof, and Philippe Schnoebelen for useful discussions.

\end{document}